# TWIST: Precision Measurement of the Muon Decay Parameters

R.E. Mischke (for the TWIST collaboration)
*TRIUMF, Vancouver, BC V6T 2A3, CANADA*

The TWIST experiment has made a precision measurement of three of the decay parameters in muon decay. The newest results are ρ = 0.75014 ±0.00017(stat) ±0.00044(sys) ±0.00011(η) and δ = 0.75067 ±0.00030(stat) ±0.00067(sys). Together with previously published results, improved constraints on possible extensions of the electroweak Standard Model are derived.

## 1. INTRODUCTION

Muon decay is an excellent laboratory for testing the electroweak Standard Model (SM). It is a purely leptonic process with the positive muon decaying into a positron and two neutrinos. The matrix element for the most general Lorentz invariant, derivative free expression [1] is described by 10 complex model-independent couplings $g_{\varepsilon m}^{\kappa}$:

$$M = \frac{4G_F}{\sqrt{2}} \sum_{\substack{\varepsilon=L,R \\ m=L,R \\ \kappa=S,V,T}} g_{\varepsilon m}^{\kappa} \langle \psi_{e_\varepsilon} | \Gamma^\kappa | \psi_{\nu_e} \rangle \langle \psi_{\nu_\mu} | \Gamma_\kappa | \psi_{\mu_m} \rangle \qquad (1)$$

where, in the SM $g_{LL}^V = 1$ and $g_{\varepsilon m}^{\kappa} = 0$ otherwise. Experimentally only the positron is measured and the decay spectrum is usually written in terms of four parameters: the Michel parameter ρ [2], δ, ξ, and η in the expression

$$\frac{d^2\Gamma}{dx d(\cos\theta)} \propto F_{IS}(x;\rho,\eta) + F_{AS}(x;\delta) P_\mu \xi \cos\theta \qquad (2)$$

where

$$F_{IS}(x) = x(1-x) + \rho \frac{2}{9}(4x^2 - 3x - x_0^2) + \eta x_0(1-x) + R.C. \qquad (3)$$

$$F_{AS}(x) = \frac{1}{3}\sqrt{x^2 - x_0^2}\left\{1 - x + \frac{2}{3}\delta\left[4x - 3 + \left(\sqrt{1-x_0^2} - 1\right)\right]\right\} + R.C. \qquad (4)$$

$x = E_e/E_{max}$, $x_0 = m_e/E_{max}$, cosθ is the angle between the positron momentum and the muon spin, and $P_\mu$ is the muon polarization.

The decay parameters can be written as bilinear combinations of the $g_{\varepsilon m}^{\kappa}$. The SM predictions are ρ = δ = 3/4, $P_\mu = \xi$ = 1, and η = 0. Precision measurements of these parameters will test the SM predictions and are sensitive to extensions to the SM. The TRIUMF Weak Interaction Symmetry Test (TWIST) experiment has made new measurements of three of these parameters resulting in improved constraints on extensions to the SM.

## 2. EXPERIMENT

The TWIST experiment used a 500 MeV proton beam incident on a graphite production target. The muons from pions that decayed on the surface of the target were transported by the M13 surface muon channel to the detector, which was located in the bore of a 2 T solenoidal magnet. Ahead of the magnet was a removable time expansion chamber [3] to measure the emittance of the muon beam. At the center of the detector was a thin Al stopping target. Positrons from muon decay were detected in an array of planar MWPC and drift chambers arranged symmetrically upstream and downstream of the stopping target.[4] Analysis of the data produced a spectrum of reconstructed energies and angles of the positrons. This spectrum was compared to a simulated spectrum to extract the decay parameters. The information for the simulated positrons was generated in the same format as the data and analyzed using the same codes. However, the simulated spectrum was generated with hidden values of the decay parameters, so the actual values for the data were not revealed until after all of the corrections and systematic uncertainties were determined.

Because it was important for the simulation to closely reproduce the experiment, validation of the physics in the simulation was essential. One important test was to stop muons at the upstream end of the detector. Then each decay positron that traversed the whole detector could be analyzed separately in the upstream and downstream halves and distributions of the energy loss and scattering in the stopping target could be determined. Comparison of these distributions for data and simulation showed excellent agreement.

## 3. SYSTEMATIC UNCERTAINTIES

The errors on this experiment are dominated by systematic uncertainties. These are estimated by exaggerating each candidate effect in the data or simulation and then scaling the sensitivity of the decay parameters by the size of the effect in the data. A summary of the systematic uncertainties for each of the parameters is presented in Table 1. The entries for $\rho$ and $\delta$ are for the latest results,[5] while those for $P_\mu\xi$ are from previously published results.[6]

Table 1: Systematic uncertainties for the decay parameters.

| units of $10^{-4}$ | $\rho$ | $\delta$ | $P_\mu\xi$ |
|---|---|---|---|
| Chamber response | 2.9 | 5.2 | 10 |
| Positron interactions | 1.6 | 0.9 | 3 |
| Alignment | 0.3 | 0.3 | 3 |
| Momentum calibration | 2.9 | 4.1 | 2 |
| Radiative corrections | <0.1 | <0.1 | 1 |
| Other | 1.1 | 0.4 | 4 |
| Fringe field depol | -- | -- | 34 |
| Stopping tgt depol | -- | -- | 12 |
| **Total** | **4.6** | **6.7** | **38** |

## 4. RESULTS

First results from TWIST have been published for $\rho$ [7] and $\delta$ [8]. The results presented here supersede those results and are based on an analysis of data taken in 2004, which were previously analyzed for the first measurement by TWIST of $P_\mu\xi$ [6]. The new results [5] are $\rho = 0.75014 \pm 0.00017(\text{stat}) \pm 0.00044(\text{sys}) \pm 0.00011(\eta)$, where the last uncertainty is due to the correlation between $\eta$ and $\rho$, and $\delta = 0.75067 \pm 0.00030(\text{stat}) \pm 0.00067(\text{sys})$. The pre-TWIST results and previously published and current results from TWIST are shown graphically in Fig. 1. The projections for the precision of the final results are also shown as TWIST expects to reach its goal of an order of magnitude improvement over previous experiments.

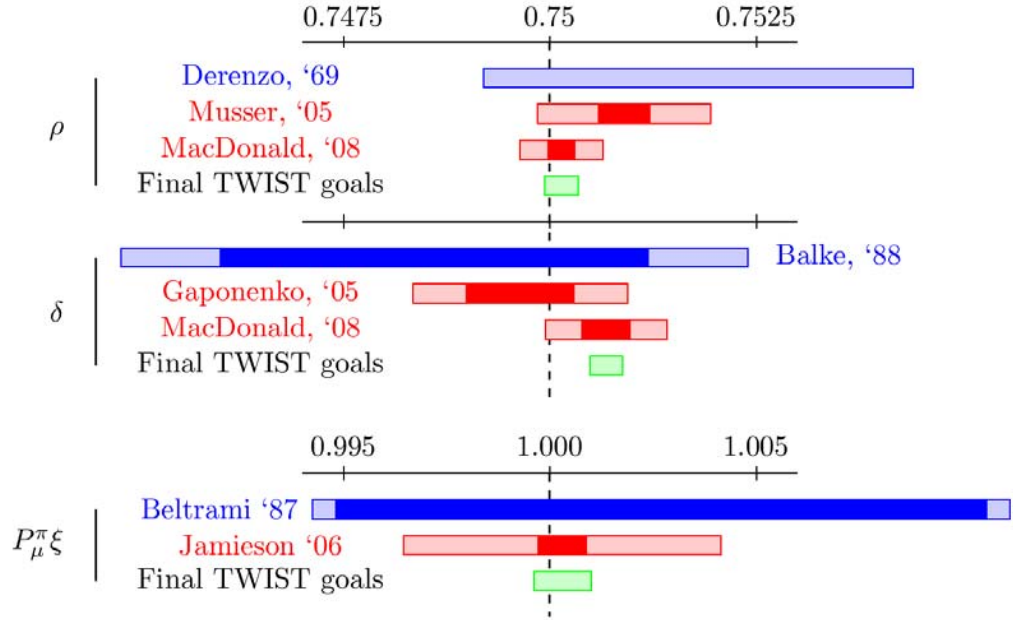

Figure 1. Graphical summary of results for the decay parameters.

### 4.1. Global analysis

TWIST is able to significantly reduce the limits for three of the general coupling constants in Eq. 1. The 90% confidence limits (C.L.) for these are shown in Table 2 along with the limits from pre-TWIST results and those from a global analysis that included the first published results from TWIST. [9]

Table 2: Global analysis of muon data

|  | pre-*TWIST* | Gagliardi et al. | MacDonald '08 |
|---|---|---|---|
| \|gSLR\| | < 0.125 | < 0.088 | **< 0.074** |
| \|gVLR\| | < 0.060 | < 0.036 | **< 0.025** |
| \|gTLR\| | < 0.036 | < 0.025 | **< 0.021** |

### 4.2. Right-handed muon decay

The quantity

$$Q_R^\mu = \frac{1}{4}\left|g_{LR}^S\right|^2 + \frac{1}{4}\left|g_{RR}^S\right|^2 + \left|g_{LR}^V\right|^2 + \left|g_{RR}^V\right|^2 + 3\left|g_{LR}^T\right|^2 \quad (5)$$

represents the probability for the decay of a right-handed muon into any type of electron and is zero in the SM. The current limit is $Q_R^\mu < 0.0024$ (90% C.L.), which is a significant improvement over the pre-TWIST limit of $Q_R^\mu < 0.005$.

### 4.1. Left-right symmetric models

Left-right symmetric models extend the SM with a right-handed W. The TWIST result for ρ provides the best constraint on the mixing angle between $W_L$ and $W_R$. The current limit is $|\zeta_g| < 0.022$ (90% C.L.), compared to the pre-TWIST limit of $|\zeta_g| < 0.066$. Coupled constraints on the mass for a right-handed W and the mixing angle are shown in Fig. 2.

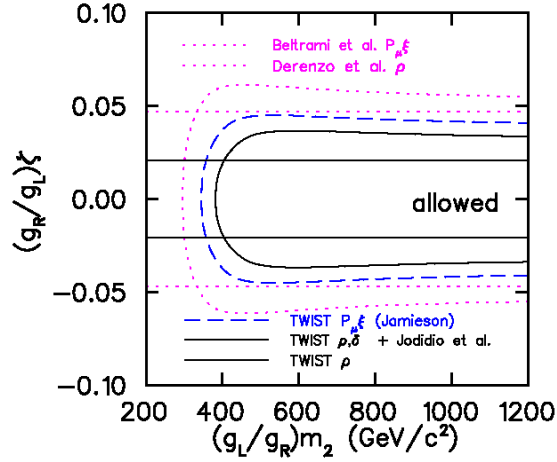

Figure 2: Constraints on left-right symmetric models from muon decay.

To date the results from TWIST are consistent with the SM; final results are anticipated within the next year.

## Acknowledgments

This work was supported in part by the Natural Sciences and Engineering Research Council of Canada, the National Research Council of Canada, the Russian Ministry of Science, and the U.S. Department of Energy. Computing resources for the analysis were provided by WestGrid.